\documentclass[twocolumn,prb,showpacs,preprintnumbers,amsmath,amssymb]{revtex4}
\usepackage{graphicx}
\usepackage{dcolumn}
\usepackage{bm}
\usepackage{amsfonts}
\usepackage{amsmath}
\usepackage{amssymb}
\begin{document}
\title{Self-consistent model of unipolar transport in organic semiconductor diodes: 
accounting for a realistic density-of-states distribution}
\author{S.V. Yampolskii}
\email{yampolsk@tgm.tu-darmstadt.de}
\author{Yu.A. Genenko}
\author{C. Melzer}
\author{H. von Seggern}  
\affiliation{Institute of Materials Science, Darmstadt University of Technology, 
Petersenstra{\ss}e 23, 64287 Darmstadt, Germany}%
\date{\today}

\begin{abstract}

A self-consistent, mean-field model of charge-carrier injection and unipolar transport in 
an organic semiconductor diode is developed utilizing the effective transport energy concept 
and taking into account a realistic density-of-states distribution as well as the presence of 
trap states in an organic material. The consequences resulting 
from the model are discussed exemplarily on the basis of an indium tin oxide/organic 
semiconductor/metallic conductor structure. A comparison of the theory to experimental data 
of a unipolar indium tin oxide/poly-3-hexyl-thiophene/Al device is presented. 

\end{abstract}


\maketitle

\section{\label{sec:intro}Introduction}

During the last years the problem of an adequate description of charge injection and transport 
in organic semiconductors (OSCs) became of great importance in view of application of these materials 
as basic elements of electronic devices such as organic field-effect transistors, organic 
light-emitting diodes, or organic photovoltaic cells~\cite{review2010}. 

Organic systems are typically wide-bandgap disordered semiconductors possessing as a rule 
a relatively narrow intrinsic density-of-states (DOS) distribution without sharp band edges typically 
approximated by a Gaussian function~\cite{Baessler1993}. Transport in such systems is generally 
described by models based on hopping of charge carriers between localized states, which are 
disordered in space and energy. In addition to the intrinsic DOS distribution, disordered OSCs 
may exhibit deeper localized states originating from impurities, or from chemical and structural 
defects. Those states are normally located energetically well below the intrinsic DOS distribution 
and, therefore, they are referred to as deep traps. In general, the energy distributions of such trap 
states are often assumed to be also of Gaussian shape~\cite{Schmechel2004}, however for some organic materials 
they are assumed to distribute exponentially~\cite{Blom2007,Mensfoort2009}. 

To simplify the description of transport, a theoretical concept is often applied where charge transport 
is controlled by carrier jumps to the so-called effective transport energy level~\cite{Monro1985,Baranovskii1997,Arkhipov1999}. This characteristic energy is localized within the intrinsic 
DOS-distribution and is defined as the energy of those sites a charge carrier visits with highest probability, 
independently of its energetic starting position. This concept allows one to reduce the description 
of hopping transport to a multiple trapping formalism~\cite{Arkhipov1982}. The transport energy plays 
then the role of the mobility edge in disordered materials~\cite{edge1} and may be used to qualitatively 
separate mobile and immobile carriers in OSCs. In this concept, charge transport is provided by 
{\it mobile} carriers with a constant mobility where all temperature dependences in the transport 
description result from the Fermi distribution of carriers~\cite{Arkhipov2001APL,Heil2003} and from 
the temperature dependence of the transport energy~\cite{Arkhipov2003}. Another version of the transport energy 
approach utilizes the effective, average mobility of the {\it whole} ensemble of carriers. In this case, the mobility reveals, in general, dependences on the temperature and on the carrier density~\cite{Baranovskii1997,Arkhipov2001,Emelianova2008}.

By now, the alternative concept of charge transport in OSCs has been also developed 
where all carriers in intrinsic DOS are considered mobile and their drift mobility is assumed to be dependent on the temperature, on the carrier density and, in general, on the electric field~\cite{Blom1996,Malliaras1,Blom2000,Tanase2004}. 
Nevertheless, as regards the last dependence, it was noted in the literature that that the field dependence of the carrier mobility is not strictly necessary, especially at high temperatures~\cite{Tanase2004,Pasveer2005}. 
The most advanced version of this approach, the so-called extended Gaussian disorder model (EGDM)~\cite{Mensfoort2009,Pasveer2005,Coehoorn2005,Mensfoort2008,Mensfoort2010}, has demonstrated good ability to simulate the current-voltage ({\it I-V}) characteristics of different OSC devices.

Another important process providing the functionality of organic devices is the injection of carriers 
from electrodes into the organic layer in view of a low intrinsic charge carrier density.
However, the description of this process remains still controversial. For charge transport in bulk OSCs the electric 
field at the injecting interface is often taken equal to zero assuming space-charge limitation of the 
current~\cite{Lampert,Heil2003,Arkhipov2001APL,Arkhipov2003,Blom1996}. 
Numerous papers assume finite values of the interfacial electric field and consider 
injection essentially as a single-particle process. Then injection is simulated as 
Fowler-Nordheim (FN) tunnelling through the surface energetic barrier~\cite{Walker2002,Malliaras1998} 
or Richardson-Schottky (RS) thermionic emission over this barrier under an applied external field~\cite{Walker2002,Arkhipov1999,Arkhipov1998,Burin2000,Preezant2003,Hosseini2005,Mensfoort2008,Mensfoort2010}. 
In fact, the field and the charge carrier density at the interface are not known because these values depend 
on the height of the injection barrier which in turn is field-dependent. Thus, for a proper description 
of the charge carrier injection, a self-consistent determination of the field and the carrier density 
at the electrode/OSC interface is necessary which would contain both limits of weak (single-particle) 
and strong (many-particle) injection including the space-charge limited regime. An attempt of such a 
self-consistent approach was undertaken in Ref.~\cite{Emtage1966} where drift-diffusion and Poisson 
equations in the bulk were considered in conjunction with an injection described in the spirit of the 
RS model. However, application of the single-particle injection as a boundary condition for the many-particle 
equations makes this approach questionable. In view of disordered OSCs, sophisticated extensions of 
the RS model with account of possible injection into the tail states below the barrier were 
developed~\cite{Arkhipov1999,Arkhipov1998,Burin2000,Arkhipov2003,Mensfoort2008,Mensfoort2010,Ruhstaller2010} 
which, however, remain essentially single-particle injection models.

Another approach to charge carrier injection including Schottky barrier lowering and space-charge effects 
was recently developed in Ref.~\cite{Holst2009} where three-dimensional (3D) hopping of charge carriers 
on sites of a cubic lattice with randomly distributed energy levels was considered as well as a 
sophisticated one-dimensional (1D) continuous model. The site occupancies and the electric field were 
calculated self-consistently by solving the three-dimensional master equation and the Poisson equation 
in successive iterations with account of the field-dependent injection barriers. Yet, 
the effect of the individual image potential was thereby overestimated because of duplication with the mean 
field deep in the sample.

Recently, a self-consistent continuous description of injection in insulating media in terms of carrier 
densities and mean fields was developed~\cite{Neumann2006,Neumann2007,Yampolskii2008,Genenko2010} based 
on the matching of the electric displacement and the electrochemical potential at the interface while still 
accounting for discreteness of injected charge carriers. The latter aspect is important for a wide range of 
values of injection barriers and applied voltages where the individual image force dominates the injection 
process. This 1D model exhibits a plausible crossover from the barrier-dominated behaviour at low voltages 
to the space-charge-dominated behaviour at high voltages and reveals a field-induced reduction of 
the injection barrier as well. The barrier lowering relates both to the Schottky effect and the voltage drop
in the electrodes initiated by substantial interfacial charge carrier transfer. 
The model applies directly to inorganic crystal insulators and wide-bandgap 
non-degenerate semiconductors as well as to very narrow-band insulators and semiconductors as was indicated 
in Ref.~\cite{Genenko2010}. It applies also to wide-bandgap OSCs if the narrow band approximation assuming 
a negligible width of the Gauss DOS is valid. The latter restriction, however, seems to fail in many organic 
semiconductors~\cite{Baessler1993,Schmechel2004,Walker2002} and excludes the possibility of injection into 
the tail states which proved to be essential for disordered 
semiconductors~\cite{Arkhipov1999,Arkhipov1998,Burin2000,Arkhipov2003,Ruhstaller2010}.

In the present paper the self-consistent approach of Refs.~\cite{Neumann2006,Neumann2007,Genenko2010} is extended  
to account for a realistic DOS shape of the organic material. The charge transport in 
semiconductor/OSC/conductor diode structure is modeled using the transport energy concept in the spirit of 
Refs.~\cite{Arkhipov2001APL,Heil2003,Arkhipov2003}  
and focusing consideration on the influence of the injection barrier heights and the DOS parameters on the {\it I-V}  
characteristics. As an example, a system is studied where only holes are injected from an indium tin oxide 
(ITO) electrode into the OSC layer, whereas the possible injection of electrons from the metallic conductor 
electrode into the OSC as well as subsequent recombination effects are excluded at this stage, for 
simplicity. The obtained model will be compared to experiments of Ref.~\cite{Heil2003} where  
{\it I-V} characteristics of poly-3-hexyl-thiophene (P3HT) based unipolar diodes were analyzed assuming space 
charge limited current boundary conditions at the electrodes.

\section{\label{sec:generalmodel}Theoretical model}

\begin{figure}[!bp]
\includegraphics[width=8cm]{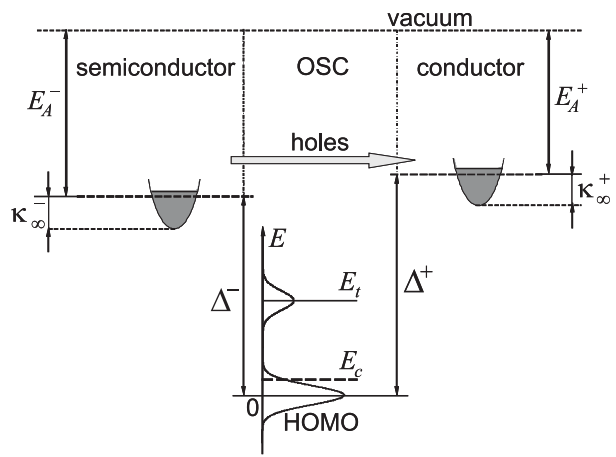}
\caption{Schematic band diagram of the considered semiconductor/OSC/conductor structure 
and the density-of-states distribution in its insulator constituent. The assumed hole transport under voltage application 
is also depicted.} \label{fig1}
\end{figure}

Let us consider an OSC layer of thickness $L$ sandwiched in between a heavily doped semiconductor and a metallic conductor 
electrode. The organic layer is supposed to be extended over the space with $-L/2\leq x\leq L/2$, whereas 
the semiconductor and conductor electrodes are extended over the half-spaces with $x<-L/2$ and $x>L/2$, 
respectively. ITO, being an electron-conducting semiconductor with a deep lying conduction band, is considered 
as the hole-injecting electrode~\cite{38,21} while Al is considered as collecting electrode. The band structure of the system under consideration is 
shown schematically in Fig.~\ref{fig1}.

The OSC is characterized by a DOS represented here as a superposition of two Gaussian DOSs (see, 
for example, Refs.~\cite{Baessler1993,Schmechel2004,Heil2003}). The first, intrinsic DOS represents the 
highest occupied molecular level (HOMO band) and the second one describes the spatially and energetically 
distributed trap states (see Fig.~\ref{fig1}). The total DOS distribution then can be written as:
\begin{equation}
\label{2Gauss}
g \left( E \right) =  \frac{P_c}{\sqrt{2 \pi}\sigma_c} \exp \left( -\frac{E^2}{2 \sigma_c^2} \right)
+ \frac{P_t}{\sqrt{2 \pi} \sigma_t} \exp \left[ -\frac{\left( E-E_t \right)^2}{2 \sigma_t^2} \right],
\end{equation}
\noindent where $E_t$ denotes the average trap energy, $\sigma_c$ and $\sigma_t$ the widths of the
intrinsic and trap DOS parts, $P_c$ and $P_t$ the numbers of intrinsic states and traps, respectively
(notice that the level $E=0$ coincides with the DOS maximum of the HOMO band).

Spatial distributions of the charge-carrier density $p$ and of the electric field $F_i$ in the OSC are 
described by the following system of equations:
\begin{equation}
kT \mu_i \frac{d p_c \left( x \right)}{d x}-e \mu_i p_c \left( x \right) F_i \left( x \right) =-j,
\label{drift-diff-eq}
\end{equation}
\begin{equation}
\frac{d F_i \left( x \right)}{d x} = \frac{e}{\epsilon_i \epsilon_0} p \left( x \right),
\label{Gauss-law}
\end{equation}
\noindent where $j$ is the position-independent steady-state current density, 
$\mu_i$ denotes the hole mobility in the conductive states of the OSC, $\epsilon_i$ 
its static relative permittivity, 
$k$ the Boltzmann constant, $e$ the positive elementary charge, $\epsilon_0$ the dielectric 
permittivity of vacuum, and $T$ the absolute temperature. 
The total density of charge carriers 
$p \left( x \right) = p_c \left( x \right) + p_t \left( x \right)$ 
is the sum of mobile and immobile carrier densities, which are defined as 
\begin{equation}
\label{p_mobile}
p_c \left( x \right) =  \int_{-\infty}^{E_c} g \left( E \right) f_p \left( E \right) d E 
\end{equation}
\noindent and
\begin{equation}
\label{p_trap}
 p_t \left( x \right)= \int_{E_c}^{\infty} g \left( E \right) f_p \left( E \right) d E,
\end{equation}
\noindent with $f_p \left( E \right) $ the Fermi-Dirac distribution of holes,
\begin{equation}
\label{Fermi}
f_p \left( E \right) = \left\{ \exp \left[ \frac{\kappa_i \left( x \right) + 
e \phi_i \left( x \right) - E}{kT}\right] + 1 \right\}^{-1}.
\end{equation}
\noindent Here, $E_c$ denotes the effective energy of the transport level, $\kappa_i$ the electrochemical 
potential for holes and $\phi_i$ the electrostatic potential in the organic layer.

Assuming $ \kappa_i \left( x \right) + e \phi_i \left( x \right) - E_c \gg kT $, 
Boltzmann statistics for the mobile carriers can be applied and one obtains
\begin{equation}
p_c \left( x \right) = N_c \exp \left[ \frac{-\kappa_i \left( x \right) - 
e \phi_i \left( x \right)}{kT}\right],  \label{pc_Boltzmann}
\end{equation}
\noindent where
\begin{align}
N_c & = \int_{-\infty}^{E_c} g \left( E \right) \exp \left(\frac{E}{kT} \right) d E \nonumber \\
& = \frac{P_c}{2} \exp \left[ \frac{\sigma_c^2}{2 \left(k T \right)^2} \right] \mbox{erfc} 
\left( -\frac{E_c}{\sqrt{2} \sigma_c} + \frac{\sigma_c}{\sqrt{2} kT} \right) \nonumber \\
& + \frac{P_t}{2} \exp \left[ \frac{E_t}{kT} + \frac{\sigma_t^2}{2 \left(k T \right)^2} \right] \mbox{erfc} 
\left( \frac{E_t-E_c}{\sqrt{2} \sigma_t} + \frac{\sigma_t}{\sqrt{2} kT} \right),
\end{align}
\noindent can be interpreted as the effective total number of states available for mobile carriers in the 
transport band of the OSC, and $\mbox{erfc} \left( z \right)$ is the complementary error function~\cite{Abramowitz}. 
Using Eq.~(\ref{pc_Boltzmann}) in combination with Eqs.~(\ref{Fermi}) and (\ref{p_trap}), we can express the 
density $p_t \left( x \right)$ through $p_c \left( x \right)$ as follows
\begin{equation}
p_t \left( x \right) = p_c \left( x \right) \int_{E_c}^{\infty} \frac{g \left( E \right) d E}
{N_c \exp \left( -E/kT \right) + p_c \left( x \right)},
\label{p-trap-pc}
\end{equation}
and, correspondingly, the total carrier density $p \left( x \right) $ as
\begin{equation}
p \left( x \right) =p_c \left( x \right)+ p_c \left( x \right) \int_{E_c}^{\infty} \frac{g \left( E \right) d E}
{N_c \exp \left( -E/kT \right) + p_c \left( x \right)}.
\label{p-total-pc}
\end{equation}
\noindent The latter equation together with Eqs.~(\ref{drift-diff-eq}) and (\ref{Gauss-law}) presents a system 
of equations describing the stationary charge transport in the OSC where the appropriate boundary conditions 
have to be applied.

The energetic differences between the equilibrium values of chemical potential in
the electrodes far away from the electrode/OSC interfaces and the
maximum of the HOMO band in the organic layer are defined as the
injection barriers $\Delta^{\pm}$ for charge carriers (from now on, the minus and plus superscripts denote the
quantities at the interfaces $x = -L/2$ and $x = L/2$, respectively).
These barriers relate to the difference between the
electrode work functions $E_A^{\pm}$ (see Fig.~\ref{fig1}), 
\begin{equation}
E_A^- + \Delta^- = E_A^+ + \Delta^+.
\label{EADelta}
\end{equation}
Assuming neither surface charge nor dipole layers at the electrode/OSC interfaces one can require continuity 
of the electrical displacement and of the electrochemical potential across the entire device. 
This continuity results in the following nonlinear boundary conditions~\cite{Genenko2010}:
\begin{align}
\label{bound-cond}
p_{c}\left( \pm \frac{L}{2} \right) & = N_c \exp\left\{ 
-\frac{\Delta^{\pm} }{kT} \mp 
\frac{el_{TF}^{\pm}}{kT}  \left[ \frac{\epsilon_i }{\epsilon_e^{\pm}} F_i \left(\pm \frac{L}{2} \right)-
\frac{j}{\gamma _e^{\pm}}\right]  
\right. \nonumber\\ 
& \left. 
+ \frac{e}{kT} \delta \phi_{Sch}^{\pm} \theta\left( 0.2 r_s^{\pm} - x_m^{\pm} \right) 
\right\},
\end{align}
\noindent where $\gamma_e^{\pm}$ denote the specific conductivities of the electrodes, $ \theta \left( z \right)$ the 
Heaviside unit step function, $r_s^{\pm} = \left[ p_s \left( \pm L/2 \right) \right]^{-1/3}$ the 
characteristic distance between charge carriers near the respective electrode, and $x_m^{\pm} = \left\{ e/16 \pi 
\epsilon_0 \epsilon_i \left[ \mp F_i \left( \pm L/2 \right) \right] \right\}^{1/2}$ the distance from 
the respective electrode to the maximum of the single-particle Schottky potential barrier\cite{Sze2007}. 
The Thomas-Fermi screening lengths in the electrodes are introduced as
\begin{equation}
\label{LTF}
l_{TF}^{\pm}=\sqrt{\frac{2\epsilon_0\epsilon_e^{\pm} \kappa_{\infty}^{\pm}}{3e^2 p_{\infty}^{\pm}}}
\end{equation}
\noindent with $\epsilon_e^{\pm}$ being the static relative permittivities of the electrodes, 
$p_{\infty }^{\pm}$ and $\kappa_{\infty}^{\pm}$ the equilibrium values of the 
carrier density  and of chemical potential in the electrodes far away from the electrode/OSC interfaces, 
the latter ones  calculated with respect to the bottom of the respective electrode conduction band (Fig.~\ref{fig1}). 
The last term in the exponent of Eq.~(\ref{bound-cond}) accounts for the discreteness of charge carriers 
and thus determines the range of the injection barriers and field values where the individual image forces dominate 
the injection process resulting in the so-called Schottky lowering of injection barriers~\cite{Sze2007},
\begin{equation}
\label{Schottky-lowering}
e \delta \phi_{Sch}^{\pm} = \sqrt{\frac{e^3}{4 \pi \epsilon_0 \epsilon_i} \left[ \mp 
F_i \left(\pm \frac{L}{2} \right) \right]}.
\end{equation} 
Details about single particle consideration in the mean field description provided here can be found in 
Ref.~\cite{Genenko2010}.
Notice that the contribution of the current $j$ in Eq.~(\ref{bound-cond}) 
can often be neglected since  it is very small in all practical cases concerning organic 
semiconductors~\cite{Neumann2006,Neumann2007}.

The nonlinear differential equations (\ref{drift-diff-eq}) and (\ref{Gauss-law}) with 
$p \left( x \right)$ from Eq.~(\ref{p-total-pc}) and with the boundary conditions (\ref{bound-cond}) have to 
be solved numerically. Knowledge about the spatial distribution of the electric field gives access to 
the voltage drop $V$ across the system for a given current density $j$, which follows
by direct integration of the field over the device thickness~\cite{Neumann2006,Yampolskii2008} and reads:
\begin{align}
V &= l_{TF}^+ \left[ \frac{\epsilon_i }{\epsilon_e^+} F_i \left( \frac{L}{2} \right)-\frac{j}{\gamma _e^+} 
\right] + l_{TF}^- \left[ \frac{\epsilon_i }{\epsilon_e^-} F_i \left( -\frac{L}{2} \right)-\frac{j}
{\gamma _e^-}\right]  \nonumber \\ &+ \int_{-L/2}^{L/2} F_i \left( x \right) d x - V_{bi},
\label{voltage}
\end{align}
\noindent where $-V_{bi}$ is the voltage drop in the case of $j=0$, {\it i.e.}, 
the built-in potential, given by the difference of the electrode's work functions,
\begin{equation}
e V_{bi} = E_A^+ - E_A^-=\Delta^- - \Delta^+ .
\end{equation}
Now, {\it I-V} characteristics of the structure under consideration can be calculated.

\section{\label{sec:experiment}Comparison with experiment and discussion}

To test the presented model, it is applied to experimental {\it I-V} 
characteristics~\cite{Heil2003} measured on a unipolar device consisting of a single 
P3HT layer of thickness $L = 125$~nm sandwiched between ITO 
and Al electrodes. It was deduced from different experiments~\cite{Heil2003,Arkhipov2001,Malm2001} 
that the total DOS in P3HT consists of the superposition of two Gaussian peaks, similarly to Eq.~(\ref{2Gauss}).

\begin{figure}[!tbp]
\includegraphics[width=8.5cm]{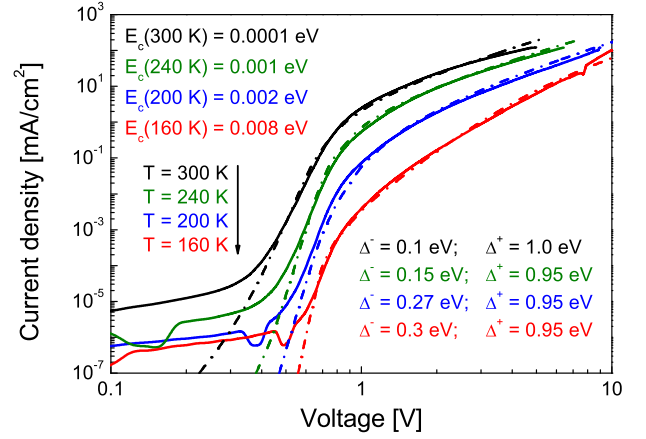}
\caption{(Color online) Measured {\it I-V} characteristics (solid curves) of ITO/P3HT/Al diode 
at different temperatures and their fitting (dashed curves) with the OSC model parameters: 
$\mu_i = 6.7 \times 10^{-4}$~cm$^2$/(V~s), $P_c = 10^{21}$~cm$^{-3}$, $P_t = 5 \times 10^{15}$~cm$^{-3}$,
$\sigma_c = 0.035$~eV, $\sigma_t = 0.02$~eV, $E_t = 0.5$~eV.} 
\label{fig2}
\end{figure}

Figure~\ref{fig2} depicts {\it I-V} characteristics of the ITO/P3HT/Al structure 
from Ref.~\cite{Heil2003} measured at different temperatures as well as best fits  
calculated with the OSC model parameters indicated. All parameters of the electrodes as well as the 
typical value of the OSC relative permittivity, $\epsilon_i =3$, are taken over from Ref.~\cite{Genenko2010}. 
At first, fitting the $T=300$~K curve, we establish the best values for DOS parameters and carrier mobility. 
Next, by fitting of the  curves for the other temperatures, these model parameters remain unchanged and only 
the barrier heights $\Delta^{\pm}$ and the transport energy $E_c$ are allowed to vary with temperature. 
One can see that the calculations reproduce satisfactorily the magnitude and the general form of the {\it I-V} characteristics in the wide range of applied voltages and at temperatures between 160 and 300~K with the fitting 
parameters indicated in the figure. Remarkable deviation of the theory from the experiment in the low-voltage parts 
of {\it I-V} characteristics (below $-V_{bi}$) is explained by a parallel leakage current which is not taken into account in the presented model.

The obtained fitting values of the transport energy are in a qualitative 
agreement with the temperature dependence of this value predicted in the literature 
(see, for example, Refs.~\cite{Baranovskii1997,Arkhipov2001}). 
However, this energy is not the only parameter needed to be varied with temperature. The best fitting can be 
achieved only if we assume the barriers $\Delta^{\pm}$ to change with temperature too. There are at least two 
reasons which could make such changes likely. One can see that the ITO/OSC barrier increases from 0.1 to 0.3~eV with 
decrease of the temperature from 300 to 160~K. Such a change corresponds qualitatively to the temperature dependence 
of the bandgap energy well-known in classical semiconductors~\cite{Sze2007} and recently established 
in charge transfer complexes based on OSC thin films~\cite{Yakuphanoglu2005,Yakuphanoglu2007} (it should be noted, 
however, that such an effect is not yet known in P3HT). On the other hand, the direction of the barrier change for 
the collecting electrode is different from that for the injecting one. It first slightly decreases and afterwards does not 
change with the temperature decrease, contrary to the increase of the injecting barrier. It may be supposed that the 
barrier heights are additionally affected by dipole layers which may emerge at the electrode/OSC interfaces, 
typically for these systems~\cite{Ishii,Kahn}. They can have an intrinsic, individual 
temperature behavior and thus may contribute to the temperature dependence of injection barriers.

It should be noted that the obtained fitting clearly emphasizes the essential role of the self-consistent 
boundary conditions in the proposed theoretical model. In Ref.~\cite{Heil2003}, experimental {\it I-V} curves 
have been simulated within the basically similar mobility edge concept but using the temperature-independent 
transport energy $E_c =0$ and the space-charge limited boundary conditions of $F_i \left(\pm L/2 \right) = 0$. 
There, the simulation has revealed less satisfactory agreement between the calculated and experimental curves 
and unreasonably low number of intrinsic states of the order of $10^{18}$~cm$^{-3}$ in contrast to the number 
of $10^{21}$~cm$^{-3}$ obtained in our simulations.

In conclusion, a one-dimensional mean-field model has been presented which describes self-consistently 
the charge-carrier transport across a semiconductor/OSC/conductor structure accounting for effects of discreteness 
of injected carriers and a realistic DOS distribution in the organic layer. The energy distribution of DOS 
was modelled by the superposition of two Gaussian peaks for the HOMO-level and deep traps. In the framework of the 
transport energy concept, assuming temperature-dependent injection barriers, good 
qualitative and quantitative agreement between the experimental and simulated {\it I-V} characteristics has been 
obtained in a wide temperature range. Compared with the popular EGDM model our approach has 
a comparable number of fitting parameters and gives similar fitting accuracy (cf., for example, with 
Refs.~\cite{Pasveer2005,Mensfoort2009,Mensfoort2010}).

\begin{acknowledgments}
Useful discussions with E.~Emelianova and V.~Nikitenko are gratefully acknowledged. This work was supported by the 
Deutsche Forschungsgemeinschaft through the Sonderforschungsbereich 595 ``Electrical Fatigue in Functional Materials".
\end{acknowledgments}

\bibliographystyle{plain}
\bibliography{apssamp}

\begin{thebibliography}{99}
\bibitem{review2010} {\it Organic electronics}, edited by G. Meller and T. Grasser, Adv. Polym. Sci. 
{\bf 223}, 1-328 (2010).
\bibitem{Baessler1993} H. B\"{a}ssler, Phys. Status Solidi B {\bf 175}, 15 (1993).

\bibitem{Schmechel2004} R. Schmechel and H. von Seggern, Phys. Status Solidi A {\bf 201}, 1215 (2004).

\bibitem{Blom2007} M.M. Mandoc, B. de Boer, G. Paasch, and P.W.M. Blom, Phys. Rev. B {\bf 75}, 193202 (2007).

\bibitem{Mensfoort2009} S.L.M. van Mensfoort, J. Billen, S.I.E. Vulto, R.A.J. Janssen, and R. Coehoorn, Phys. Rev. B {\bf 80}, 033202 (2009).

\bibitem{Monro1985} D. Monroe, Phys. Rev. Lett. {\bf 54}, 146 (1985).

\bibitem{Baranovskii1997} S.D. Baranovskii, T. Faber, F. Hensel, and P. Thomas, 
J. Phys.: Condens. Matter {\bf 9}, 2699 (1997).

\bibitem{Arkhipov1999} V.I. Arkhipov, U. Wolf, and H. B\"{a}ssler, Phys. Rev. B {\bf 59}, 7514 (1999).

\bibitem{Arkhipov1982} V.I. Arkhipov and A.I. Rudenko, Phil. Mag. B {\bf 45}, 189 (1982). 

\bibitem{edge1} N.F. Mott, {\it Conduction in Non-Crystalline Materials} (Clarendon, Oxford, 1987).

\bibitem{Arkhipov2001APL} V.I. Arkhipov, P. Heremans, E.V. Emelianova, and G.J. Adriaenssens, Appl. Phys. Lett. {\bf 79},  4154 (2001).

\bibitem{Heil2003} V.R. Nikitenko, H. Heil, and H. von Seggern, J. Appl. Phys. {\bf 94}, 2480 (2003).

\bibitem{Arkhipov2003} V.I. Arkhipov, H. von Seggern, and E.V. Emelianova, Appl. Phys. Lett. {\bf 83}, 5074 (2003).

\bibitem{Arkhipov2001} V.I. Arkhipov, E.V. Emelianova, and G.J. Adriaenssens, Phys. Rev. B {\bf 64}, 125125 (2001).

\bibitem{Emelianova2008} E.V. Emelianova, M. van der Auweraer, G.J. Adriaenssens, and A. Stesmans, Org. Electr. {\bf 9}, 129  (2008).

\bibitem{Blom1996} P.W.M. Blom, M.J.M. de Jong, and J.J.M. Vleggaar, Appl. Phys. Lett. 
{\bf 68}, 3308 (1996).

\bibitem{Malliaras1} G.G. Malliaras and J.C. Scott, J. Appl. Phys. {\bf 85}, 7426 (1999).

\bibitem{Blom2000} P.W.M. Blom and M.C.J.M. Vissenberg, Mater. Sci. Eng. R-Rep. {\bf 27}, 53 (2000).

\bibitem{Tanase2004} C. Tanase, P.W.M. Blom, and D.M. de Leeuw, Phys. Rev. B {\bf 70}, 193202 (2004).

\bibitem{Pasveer2005} W.F. Pasveer, J. Cottaar, C. Tanase, R. Coehoorn, P.A. Bobbert, P.W.M. Blom, D.M. de Leeuw, and 
M.A.J. Michels, Phys. Rev. Lett. {\bf 94}, 206601 (2005).

\bibitem{Coehoorn2005} R. Coehoorn, W.F. Pasveer, P.A. Bobbert, and M.A.J. Michels, Phys. Rev. B {\bf 72},155206 (2005).

\bibitem{Mensfoort2008} S.L.M. van Mensfoort and R. Coehoorn, Phys. Rev. B {\bf 78}, 085207 (2008).

\bibitem{Mensfoort2010} S.L.M. van Mensfoort, V. Shabro, R.J. de Vries, R.A.J. Janssen, and R. Coehoorn, J. Appl. Phys. {\bf 107}, 113710 (2010).

\bibitem{Lampert} M.A. Lampert and P. Mark, {\it Current injection in solids} (Academic Press, New York, 1970).

\bibitem{Malliaras1998} G.G. Malliaras and J.C. Scott, J. Appl. Phys. {\bf 83}, 5399 (1998).

\bibitem{Walker2002} A.B. Walker, A. Kambili, and S.J. Martin, J. Phys.: Condens. Matter {\bf 14}, 9825 (2002).

\bibitem{Arkhipov1998} V.I. Arkhipov, E.V. Emelianova, Y.H. Tak, and H. B\"{a}ssler, J. Appl. Phys. {\bf 84}, 848 (1998).

\bibitem{Burin2000} A.L. Burin and M.A. Ratner, J. Chem. Phys. {\bf 113}, 3941 (2000).

\bibitem{Preezant2003} Y. Preezant and N. Tessler, J. Appl. Phys. {\bf 93}, 2059 (2003).

\bibitem{Hosseini2005} A.R. Hosseini, M.H. Wong, Y. Shen, and G.G. Malliaras, J. Appl. Phys. {\bf 97}, 023705 (2005).

\bibitem{Emtage1966} P.R. Emtage and J.J. O'Dwyer, Phys. Rev. Lett. {\bf 16}, 356 (1966).

\bibitem{Ruhstaller2010} E. Knapp, R. H\"{a}usemann, H.U. Schwarzenbach, and B. Ruhstaller, J. Appl. Phys. {\bf 108}, 054504  (2010). 

\bibitem{Holst2009} J.J.M. van der Holst, M.A. Uijttewaal, B. Ramachandhran, R. Coehoorn, P.A. Bobbert, G.A. de Wijs, 
and R.A. de Groot, Phys. Rev. B {\bf 79}, 085203 (2009).

\bibitem{Neumann2006} F. Neumann, Y.A. Genenko, C. Melzer, and H. von Seggern, J. Appl. Phys. {\bf 100}, 084511 (2006).

\bibitem{Neumann2007} F. Neumann, Y.A. Genenko, C. Melzer, S.V. Yampolskii, and H. von Seggern, Phys. Rev. B {\bf 75}, 205322  (2007).

\bibitem{Yampolskii2008} S.V. Yampolskii, Yu.A. Genenko, C. Melzer, K. Stegmaier, and H. von Seggern, J. Appl. Phys. 
{\bf 104}, 073719 (2008).

\bibitem{Genenko2010} Yu.A. Genenko, S.V. Yampolskii, C. Melzer, K. Stegmaier, and H. von Seggern, Phys. Rev. B {\bf 81}, 125310  (2010).

\bibitem{38} R. Bel Hadj Tahar, T. Ban, Y. Ohya, and Y. Takahashi, J. Appl. Phys. {\bf 83}, 2631 (1998).

\bibitem{21} S.J. Martin, A.B. Walker, A.J. Campbell, and D.C. Bradley, J. Appl. Phys. {\bf 98}, 063709 (2005).

\bibitem{Abramowitz}{\it Handbook on Mathematical Functions}, edited by 
M. Abramovitz and I. Stegun, (Dover Publications, New York, 1970).

\bibitem{Sze2007} S.M. Sze and K.K. Ng, {\it Physics of Semiconductor Devices} (Wiley, Hoboken, NJ, 2007).

\bibitem{Malm2001} N. von Malm, R. Schmechel, and H. von Seggern, Synth. Met. {\bf 126}, 87 (2001).

\bibitem{Yakuphanoglu2005} F. Yakuphanoglu, M. Arslan, M. K\"u\c{c}\"ukislamo\u{g}lu, and M. Zengin, Sol. Energy {\bf 79}, 
96 (2005).

\bibitem{Yakuphanoglu2007} F. Yakuphanoglu and M. Arslan, Physica B {\bf 393}, 304 (2007). 

\bibitem{Ishii} H. Ishii, K. Sugiyama, E. Ito, and K. Seki, Adv. Mater. (Weinheim, Ger.) {\bf 11}, 605 (1999).

\bibitem{Kahn} A. Kahn, N. Koch, and W. Gao, J. Polym. Sci., Part B: Polym. Phys. {\bf 41}, 2529 (2003).
\end{thebibliography}

\end{document}